\documentclass[12pt]{article}

\newcommand{\e}{\varepsilon}
\newcommand{\p}{\bot}
\newcommand{\dd}{\partial}
\newcommand{\de}{\delta}
\newcommand{\m}{\mu}
\newcommand{\n}{\nu}
\newcommand{\om}{\omega}
\newcommand{\Om}{\Omega}
\newcommand{\te}{\theta}
\newcommand{\et}{\eta}
\newcommand{\La}{\Lambda}
\newcommand{\ls}{\left(}
\newcommand{\lks}{\left[}
\newcommand{\rs}{\right)}
\newcommand{\rks}{\right]}
\newcommand{\ra}{\rangle}
\newcommand{\str}[1]{\mathrel{\mathop{\longrightarrow}\limits_{#1}}}
\newcommand{\stv}[1]{\mathrel{\mathop{\stackrel{#1}{\longrightarrow}}}}
\newcommand{\lo}{\longrightarrow}
\newcommand{\f}{\varphi}
\newcommand{\fc}{\check\varphi}
\newcommand{\ti}{\tilde}
\newcommand{\xt}{\tilde x}
\newcommand{\pt}{\tilde p}
\newcommand{\la}{\lambda}
\newcommand{\Pc}{{\cal P}}
\newcommand{\re}{{\rm Re}\,}
\newcommand{\im}{{\rm Im}\,}

\newcommand{\disn}[2]{$$\displaylines{\refstepcounter{equation}%
            \label{#1}\hskip 1em minus 1em #2\hfilneg}$$}
\newcommand{\nom}{\hfil\hskip 1em minus 1em (\theequation)}
\newcommand{\no}{\hfil \hskip 1em minus 1em\phantom{(\theequation)}%
            \hfilneg\cr\hfilneg\hskip 1em minus 1em\hfil}
\newcommand{\ns}{\hfill\cr\hfill}

\makeatletter

\newcount\dobav
\newcount\otlog
\newcount\vrem
\def\@citex[#1]#2{\if@filesw\immediate\write\@auxout{\string\citation{#2}}\fi
  \let\@citea\@empty
  \dobav=-1
  \otlog=-1
  \@cite{\@for\@citeb:=#2\do
    {\def\@tempa##1##2\@nil{\edef\@citeb{\if##1\space##2\else##1##2\fi}}%
     \expandafter\@tempa\@citeb\@nil
     \@ifundefined{b@\@citeb}{\@warning%
       {Citation `\@citeb' on page \thepage \space undefined}%
       \vrem=-1}{\vrem=\csname b@\@citeb\endcsname}
\advance\vrem by -1 \ifnum \vrem=\dobav
 \otlog=\vrem
 \advance\otlog by 1
\else
 \ifnum \vrem=\otlog
  \advance\otlog by 1
 \else
  \ifnum \otlog>0
   \advance\dobav by 1
   \ifnum \otlog=\dobav
    \hbox{,\penalty\@m\ \the\otlog}%
   \else
    \hbox{--\the\otlog}%
   \fi
   \otlog=-1
  \fi
  \dobav=\vrem
  \advance\dobav by 1
  \@citea\def\@citea{,\penalty\@m\ }%
  \ifnum \dobav=-1
   {\reset@font\bf ?}%
  \else
   \hbox{\the\dobav}%
  \fi
 \fi
\fi
}%
\ifnum \otlog>0
 \advance\dobav by 1
 \ifnum \otlog=\dobav
  \hbox{,\penalty\@m\ \the\otlog}%
 \else
  \hbox{--\the\otlog}%
 \fi
\fi }{#1}}

\renewcommand{\section}{\@startsection{section}{1}{0pt}%
          {3.5ex plus 1ex minus .2ex}{2.3ex plus .2ex}{\noindent\hfil\bf}}

\makeatother

\begin{document}

\title{
Quantum Fields on the Light Front,\\
Formulation in Coordinates close to the Light Front,\\
Lattice Approximation\\}

\author{
E.-M. Ilgenfritz\thanks{
Institut fur Physik, Humboldt-Universitat zu Berlin, Berlin, Germany,
e-mail: ilgenfri@physik.hu-berlin.de.},
S.~A.~Paston\thanks{
St.~Petersburg State University, St.~Petersburg, Russia,
e-mail: paston@pobox.spbu.ru.},
H.-J.~Pirner\thanks{
Institut fur Theoretische Physik, Universitat Heidelberg, Heidelberg, Germany,
e-mail: pir@tphys.uni-heidelberg.de.},\\
E.~V.~Prokhvatilov\thanks{
St.~Petersburg State University, St.~Petersburg, Russia,
e-mail: Evgeni.Prokhvat@pobox.spbu.ru.},
V.~A.~Franke\thanks{
St.~Petersburg State University, St.~Petersburg, Russia,
e-mail: franke@pobox.spbu.ru.}
}

\date{\vskip 15mm}

\maketitle

\begin{abstract}
We review the fundamental ideas of quantizing a theory on a Light Front
including the Hamiltonian approach to the problem of bound states on the
Light Front and the limiting transition from formulating a theory in
Lorentzian coordinates (where the quantization occurs on spacelike
hyperplanes) to the theory on the Light Front, which demonstrates the
equivalence of these variants of the theory. We describe attempts to find
such a form of the limiting transition for gauge theories on the Wilson
lattice.
\end{abstract}

\newpage
\section{Introduction}
The idea of quantizing relativistic fields on the Light Front (LF) was
proposed by Dirac \cite{dir}, who introduced the LF coordinates
 \disn{1}{
x^{\pm}=(x^0\pm x^3)/\sqrt{2},
\qquad x^{\p}\equiv x^k,\quad k=1,2.
\nom}
instead of the Lorentzian coordinates $x^0,x^1,x^2,x^3$.
Here, $x^+$ plays the role of time, $x^-$ (the "lightlike" coordinate)
plays the role of one of the spatial coordinates, and $x^k$ are "transverse"
coordinates, $k =1,2$.

The field theory is quantized on the hyperplane $x^+=0$, which is
tangent to the light cone and therefore corresponds to the LF.
The role of the Hamiltonian is here played by the generator
of translations along
$x^+$, i.~e., the operator $P_+=(P_0+P_3)/\sqrt{2}$, and the role
of one of the momentum components (the "lightlike"
component) is played by the generator of translations along $x^-$,
i.~e., the operator $P_-=(P_0-P_3)/\sqrt{2}$.

One advantage of quantizing on the LF is the formal simplification
of the problem of describing the quantum vacuum state in field theory.
Standardly, fields are quantized at a fixed time in the Lorentzian
coordinates, for example, at $x^0=0$, and their Fourier transforms
are merely related to "bare" creation and annihilation operators
$a^+(p)$ and $a(p)$. A "bare" (or "mathematical") vacuum is determined
by the condition
 \disn{2}{
a(p)|0\ra=0.
\nom}
Such a state corresponds to the free theory vacuum and does not
coincide with the physical vacuum of the interacting theory.
When solving the stationary Schroedinger equation in the Fock space
over this
mathematical vacuum, we must also describe the physical vacuum
state in terms of the "bare" vacuum or "bare" creation operators.
Such a description is possible in principle if we introduce ultraviolet
and infrared regularizations. But this description is immensely complicated
(authors often confine themselves to the "Gaussian" approximation
in the simplest models \cite{stev}).

For quantization in the LF coordinates, it is essential that the
lightlike component of the momentum $P_-$ be nonnegative, and $P_->0$
for states with positive squared mass. If massless physical particles
are not present, then the state with the momentum $p_-=0$ formally
describes the physical vacuum, i.~e., it also corresponds to the minimum
of the operator $P_+$. Introducing the "bare" creation and
annihilation operators on the LF, we can use them to define
the corresponding mathematical vacuum. By virtue of the structure of the
momentum operator $P_-$, the mathematical vacuum corresponds
to the minimum of this operator, i.~e., it coincides with the physical
vacuum (still formally because there are divergences in the theory).
We demonstrate this below in the example of the scalar field theory.

The Fock space constructed over this mathematical vacuum on the LF
can be used to describe solutions of the corresponding analogue
of the stationary Schroedinger equation determining the masses $M$
of bound states at fixed values of the momenta $P_-$ and $P_\p\equiv(P_1,P_2)$,
 \disn{3}{
P_+ |\psi\ra= H |\psi\ra= p_+|\psi\ra,\no
P_- |\psi\ra =p_-|\psi\ra,\no
P_{\p}|\psi\ra= 0,
\nom}
where $M^2 = 2p_+p_-$.

As an example, we consider the theory of a scalar field
(with the mass $m$) governed by the Lagrangian density $L(x)$:
 \disn{4}{
L =\frac{1}{2}\dd_\m\f\dd^\m\f-\frac{1}{2}m^2\f^2-U(\f)=
\dd_+\f\dd_-\f-\frac{1}{2}\dd_k\f\dd_k\f-\frac{1}{2}m^2\f^2-U(\f),\quad
\dd_\m\f\equiv \frac{\dd}{\dd x^\m}\f.
\nom}
To construct the canonical formalism at $x^+ =0$, we introduce
the Fourier transform of the field $\f(x)$
w.r.t. the coordinate $x^-$, taking the nonnegativity of
the momentum $p_-$ into account:
 \disn{5}{
\f(x^-,x^\p)=\frac{1}{\sqrt{2\pi}}\int\limits_0^\infty dp_-
(2p_-)^{-1/2}\ls a^+(p_-,x^\p)e^{ip_-x^-}+a(p_-,x^\p)e^{-ip_-x^-}\rs.
\nom}
Then
 \disn{6}{
\int dx^- \dd_+\f\dd_-\f=
\int\limits_0^\infty dp_-\frac{1}{2i}
\ls \dd_+a^+(p_-,x^\p)a(p_-,x^\p)-a^+(p_-,x^\p)\dd_+a(p_-,x^\p)\rs.
\nom}
which is the canonical form in which $a(p_-,x^\p)$ and $ia^+(p_-,x^\p)$
are canonically conjugate variables. The quantum operators $a(p_-,x^\p)$ and
$a^+(p_-,x^\p)$ satisfy the commutation relations
 \disn{7}{
[a(p_-,x^\p),a^+(q_-,y^\p)]=\de(p_--q_-)\de^2(x^\p-y^\p),\no
[a(p_-,x^\p),a(q_-,y^\p)]=[a^+(p_-,x^\p),a^+(q_-,y^\p)]=0.
\nom}
(at $x^+ =0$). Using the expression for the energy-momentum tensor
 \disn{8}{
T_{\m\n}=\dd_\m\f\dd_\n\f-g_{\m\n}L,
\nom}
we can find the operator $P_-$,
 \disn{8.1}{
P_-=\int d^2x^\p\int dx^-\, T_{--}=
\int d^2x^\p\int dx^-\ls \dd_-\f\rs^2=
\int d^2x^\p\int\limits_0^\infty dp_-\,\, p_-a^+(p_-,x^\p)a(p_-,x^\p),
\nom}
where we drop an infinite constant. We define the mathematical vacuum
state $|0\ra$ the conditions
 \disn{9}{
a(p_-,x^\p)|0\ra=0,\quad p_-\ge 0.
\nom}
Then $P_-|0\ra=0$, and the state $|0\ra$ can be interpreted as
the physical vacuum. For this to hold, we drop the infinite constant
in Eq.~(\ref{8.1}).

The Hamiltonian can be obtained standardly from the Lagrangian
written in terms of the variables $a(p_-,x^\p)$ and $a^+(p_-,x^\p)$.
In the theory under consideration, it coincides with the expression
 \disn{10}{
P_+=\int d^2x^\p\int dx^-\, T_{-+}=
\int d^2x^\p\int dx^-\ls
\frac{1}{2}\dd_k\f\dd_k\f+\frac{1}{2}m^2\f^2+U(\f)\rs.
\nom}
(up to a constant).

We now briefly formulate the difficulties in quantizing on the LF.

{\bf 1.}
Ultraviolet singularities in the quantum field theory require
introducing a regularization and subsequently renormalizing the theory.
In the LF quantization, it is difficult to introduce a regularization
preserving the Lorentz and gauge symmetries. Noninvariant
regularizations make the renormalization procedure more difficult and,
in particular, result in introducing unusual counterterms and relating
new arbitrary constants to them. When using numerical methods to solve
the Schroedinger equation on the LF nonperturbatively, we must work
with a regularized theory and fit constants such that the calculation
results depend only weakly on the regularization parameters.

{\bf 2.} The quantization on the LF is related to special features
and divergences when the momentum $p_-$ of "bare" quanta tends to
zero. From the standpoint of Lorentzian coordinates, these
divergences can be interpreted as ultraviolet ones as the limit $p_- \to 0$
is reached in the domain $p_3 \to \infty$ under the condition $p^2=m^2$.
But if we regularize the theory by cutting off the momentum
$p_-$ ($p_-\ge\e>0$), then we also exclude vacuum effects.
The physical vacuum
then coincides with the "bare" vacuum, which forbids condensates
and spontaneous vacuum symmetry breaking.

All the vacuum effects must be taken into account by additional
terms in the Hamiltonian on the LF. Obtaining these terms is diffcult.
For instance, their role can be played by the counterterms that
renormalize singularities as $p_- \to 0$ and reconstruct
the results of the Lorentz-covariant perturbation theory w.r.t.
the interaction constant in all orders \cite{tmf97,tmf99,tmf02}.
We can introduce these terms based on semiphenomenological considerations,
for instance, by studying the limiting transition on the LF starting
from the theory quantized on a spacelike surface close to the LF
(see \cite{naus,pred96}). We describe this method below.

There were also attempts to take vacuum effects into account
when quantizing gauge theories on the LF by analyzing the exact
operatorial solution of the Schwinger model \cite{mcdal}.

{\bf 3.}
When calculating the mass spectra of bound states, we must restrict
the consideration to a finite number of degrees of freedom of
quantum fields. For this, we can restrict the coordinates together
with introducing periodic boundary conditions on the fields in these
coordinates, $|x^-|\le L$, ${x^\p}^2\le L_\p^2$,
and also restrict the obtained discrete set of momenta.
The canonical formalism for such a formulation contains
involved constraints because there are zero Fourier modes of fields
w.r.t. the coordinate $x^-$. For gauge fields on the LF,
this problem was considered in \cite{nov2,nov2a}.
Solving the Schroedinger equation in the Fock space on the LF
with a fixed total momentum of bound states
($p_- = p_n=(\pi n)/L$, $p_{\p}=0$), with the condition $p_n > 0$,
and with the introduced cutoff over transverse momenta of separate
excitations, we obtain finite-dimensional subspaces of the Fock
space depending on $n$ for each integer $n$ (and for the restricted
set of transverse momenta). Bound-state masses can be obtained
as the limit of values of $m_n^2 = 2p_{n,+}p_{n,-}$ found in each
of the subspaces as $n \to\infty$. For a number
of (1+1)-dimensional models, the values of $m_n^2$ already become
practically independent of $n$ at not too large $n$ \cite{ann,yf05,pauli}.
But for (3+1)-dimensional theories, the dimensions of the Fock subspaces
depend on the whole set of momenta $p_{\perp}$, which makes
the calculations considerably more involved \cite{ann2}. Moreover,
this regularization breaks the Lorentz (and gauge) symmetry,
complicating the renormalization procedure and the restoration
of symmetry in the limit of removed regularization.

Another regularization method is to restrict the number of "bare"
quanta participating in constructing the bound-state wave function
in the Fock space on the LF, i.e., to introduce a "cutoff" in the Fock
space w.r.t. the total number of "particles" (the Tamm-Dankoff
method on the LF) \cite{wils,mymc}. We can then leave coordinates
unrestricted. But passing to the Fock subspace then encounters
serious technical troubles when the number of "particles"
(i.e., bare quanta) increases.

\section{The method of the limiting transition to the Hamiltonian on the LF}
Quantizing a theory on the LF can be interpreted as the formal limit
of quantization in a Lorentzian reference frame moving with a speed
close to the speed of light w.r.t. desired bound states.
We parameterize the corresponding Lorentz coordinate transformation
$x \to x'$ with the parameter $\eta>0$, $\eta \to 0$:
 \disn{11}{
x'^+ =\frac{\sqrt{2}}{\eta}\,x^+,\qquad
x'^- = \frac{\eta}{\sqrt{2}}\, x^-,\qquad
x'^{\p}=x^{\p},
\nom}
where $x'^{\pm}=(x'^0\pm x'^3)/\sqrt{2}$.
The plane of the field quantization $x'^0 =0$ approximates
the LF plane as $\eta \to 0$.

For the further discussion, we pass from the "fast moving" Lorentzian
reference frame $x'$ to the convenient coordinates $\tilde x$:
 \disn{12}{
\tilde x^+ = \eta x'^0 =x^+ + \frac{\eta^2}{2}x^-,\qquad
\tilde x^- = \eta^{-1}(x'^0 - x'^3)=x^-,\qquad
\tilde x^{\perp} = x^{\perp}.
\nom}
The plane $\tilde x^+ = 0$ then coincides with the plane $x'^0 =0$,
and the coordinates $\tilde x$ become the LF coordinates as $\eta \to 0$.
The metric tensor corresponding to these coordinates has the nonzero
components
 \disn{13}{
\ti g_{+-}(\xt)=\ti g_{-+}(\xt) =1,\qquad
\ti g_{--}(\xt)=-\eta^2,\qquad
\ti g_{kk}(\xt)=-1,\no
\ti g^{+-}(\xt)=\ti g^{-+}(\xt) =1,\qquad
\ti g^{++}(\xt)=\eta^2,\qquad
\ti g^{kk}(\xt)=-1.
\nom}
We now demonstrate the limiting-transition method in the example
of the scalar field theory written in these coordinates.
We define the Lagrangian density as
 \disn{14}{
L(\xt)=\sqrt{\ti g(\xt)}\ls\frac{1}{2}\dd_\m\f(\xt)\dd_\n\f(\xt)
\ti g^{\m\n}(\xt)-\frac{1}{2}m^2\f^2(\xt)-\la\f^4(\xt)\rs=\hfill\no\hfill=
\ti \dd_+\f(\xt)\ti \dd_-\f(\xt)+\frac{\et^2}{2}
\ls\ti\dd_+\f(\xt)\rs^2-\frac{1}{2}\ls\dd_\p\f(\xt)\rs^2-
\frac{1}{2}m^2\f^2(\xt)-\la\f^4(\xt),
\nom}
where $\la$ is the coupling constant. We define the canonical
variables at $\xt^+=0$ as
 \disn{15}{
\f(\xt),\qquad
\Pi(\xt)=\frac{\de L}{\de(\ti\dd_+\f)}=\et^2\ti\dd_+\f+\ti\dd_-\f.
\nom}
Then the Hamiltonian is
 \disn{16}{
H(\eta)=\int d^2x^\p\int d\xt^-\ls
\frac{\ls\Pi-\ti\dd_-\f\rs^2}{2\et^2}+
\frac{1}{2}\ls\dd_\p\f\rs^2+\frac{1}{2}m^2\f^2+\la\f^4\rs.
\nom}
We define the "bare" creation and annihilation operators using
the Fourier transformation,
 \disn{17}{
\f(\xt)=(2\pi)^{-3/2}\int d^2\pt\, d\pt_-\frac{1}{\sqrt{2\om_p}}
\lks a(\pt)+a^+(-\pt)\rks e^{-i\pt\xt},\no
\Pi(\xt)=(2\pi)^{-3/2}\int d^2\pt\, d\pt_-\sqrt{\frac{\om_p}{2}}
\lks a(\pt)-a^+(-\pt)\rks e^{-i\pt\xt}.
\nom}
where
$\pt\xt\equiv p_\p x^\p+\pt_-\xt^-$ and $\om_p=\ls\pt_-^2+\et^2(p_\p^2+m^2)\rs^{1/2}$.
At $\xt^+=0$, these operators satisfy the commutation relations
 \disn{18}{
[a(\pt),a^+(\ti q)]=\de^3(\pt-\ti q),\qquad
[a(\pt),a(\ti q)]=[a^+(\pt),a^+(\ti q)]=0.
\nom}
The free part of the Hamiltonian can be written in the form
 \disn{19}{
H_0=\int d^2p_\p \, d\pt_-\frac{\om_p-\pt_-}{\et^2}\,a^+(\pt)a(\pt).
\nom}
If we restrict the field modes in $\tilde p_-$ by the cutoff $|\tilde p_-|\ge\e$
and assume that the interaction Hamiltonian is normally ordered
(which was done in (\ref{19})), then we can easily see that in the limit
$\et\to 0$ under the finite-energy condition, we obtain the canonical
formulation of the theory on the LF on the subspace ${|f_0\ra}$ of the Fock
space determined by the conditions
 \disn{20}{
a(\tilde p)|f_0\ra=0,\quad \tilde p_-\le -\e.
\nom}
In the limit, this subspace becomes the Fock space on the LF.
To take the vacuum effects into account approximately, we also consider
the vicinity of $\tilde p_- \to 0$, including the field modes
with $|\tilde p_-|\le \Lambda \eta$, where $\La$ is some quantity
with the dimension of momentum. At $\eta \to 0$, we neglect field
modes in the interval $\Lambda\eta <|\tilde p_-|<\e$. We assume
that such a distortion of the theory is inessential in the limit $\eta \to 0$
if we set $\Lambda \to \infty$ upon passing to this limit
(after the corresponding ultraviolet renormalization of the theory).

We now perform the limiting transition in the framework of
the perturbation theory in the small parameter $\eta$ for the solutions
of the Schroedinger equation with Hamiltonian (\ref{16}),
 \disn{20.1}{
\ti H(\et)|f(\et)\ra=\ti E(\et)|f(\et)\ra,
\nom}
describing states with finite energy (and mass). For this, we expand
the Hamiltonian in powers of $\eta$:
 \disn{21}{
\ti H(\et)=\frac{1}{\et^2}H^{(0)}+\frac{1}{\et}H^{(1)}+H^{(2)}+\dots.
\nom}
The terms of this expansion can be obtained by substituting the field
expansions corresponding to the above mode splitting
 \disn{22}{
\f(\xt)\approx \fc_\e(\xt)+\f_{\La\et}(\xt),
\nom}
in the Hamiltonian, where the term $\fc_{\e}(\xt)$
contains only modes with $|\pt_-|\ge\e$ and $\f_{\La\et}(\xt)$
contains those with $|\pt_-|\le\La\et$.
We assume the analogous expansion for $\Pi(\tilde x)$.

The free part (quadratic in fields) of Hamiltonian (\ref{21}) can be
written in the form
 \disn{23}{
H_0(\Pi,\f)\approx H_0(\check\Pi_\e,\fc_\e)+H_0(\Pi_{\La\et},\f_{\La\et}),
\nom}
and the interacting part is
 \disn{24}{
H_I=\la\int d^2 x^\p d\xt^- \f^4(\xt)\approx
H_I(\fc_\e)+H_I(\f_{\La\et})+
\la\int d^2x^\p d\xt^-\ls 6\f_{\La\et}^2 \fc_\e^2+
4\f_{\La\et} \fc_\e^3\rs.
\nom}
We neglect the term $\f_{\La\et}^3\fc_\e$ in the integrand because
in the desired limit $\et\to 0$ the quantity $\f_{\La\et}^3$ becomes
constant and the integral over $\xt^-$ of $\fc_\e$ therefore vanishes.

Representing the fields in terms of the creation and annihilation
operators, we obtain
 \disn{25}{
H_0(\check\Pi_{\e},\fc_{\e})=
\frac{1}{\et^2}\int d^2 p_\p \int\limits_{|\pt_-|\ge\e}\!d\pt_-
\lks \sqrt{\pt_-^2+\et^2(p_\p^2+m^2)}-\pt_-\rks a^+(\pt)a(\pt)=\ns=
\frac{2}{\et^2}\int d^2 p_\p \int\limits_{p_-\le-\e}\!dp_-
\,|p_-|\,a^+(p)a(p)+\int d^2 p_\p \int\limits_{p_-\ge\e}\!dp_-
\frac{m^2+p_\p^2}{2p_-}\,a^+(p)a(p)+O(\et^2).
\nom}
On the other hand, the fields $\f_{\Lambda \eta}, \Pi_{\Lambda \eta}$
can be represented in a "fast moving" Lorentzian reference frame
in the form
 \disn{26}{
\f_{\La\et}(\xt)\Bigr|_{\xt^+=0}=\f_{\La}(-x'^3,x^\p)\Bigr|_{x'^0=0},\no
\Pi_{\La\et}(\xt)\Bigr|_{\xt^+=0}=\et\Pi_{\La}(-x'^3,x^\p)\Bigr|_{x'^0=0},
\nom}
where $\La$ is the momentum cutoff, $|p'_3|\le\La$. Hence, it is easy
to obtain
 \disn{27}{
\ti H(\Pi_{\La\et},\f_{\La\et})=
H_0(\Pi_{\La\et},\f_{\La\et})+H_I(\f_{\La\et})=
\frac{1}{\et}\ls P_0+P_3\rs_{\La,x'^0=0},
\nom}
up to a possible additive constant that makes the quantum operator
in the r.h.s. of the equality nonnegative and its minimum equal to zero.
As a result, we obtain
 \disn{28}{
H^{(0)} =2
\int d^2 p_\p \!\int\limits_{p_-\le-\e}\!dp_-
|p_-|\,a^+(p)a(p),
\nom}
 \disn{29}{
H^{(1)}=\ls P_0+P_3\rs_{\La,x'^0=0},
\nom}
 \disn{30}{
H^{(2)}=\int d^2 p_\p \int\limits_{p_-\ge\e}\!dp_-
\frac{m^2+p_\p^2}{2p_-}\,a^+(p)a(p)+\ns+
\la\int d^2x^\p d x^-\ls 6\f_{\La}^2(x)\Bigr|_{x'^0=0} \fc_\e^2(x)+
4\f_{\La}(x)\Bigr|_{x'^0=0} \fc_\e^3(x)+\fc_\e^4(x)\rs.
\nom}

We now construct the perturbation theory for Schroedinger equation
(\ref{20.1}), introducing expansions in powers of $\et$,
 \disn{31}{
|f(\et)\ra=|f^{(0)}\ra+\et |f^{(1)}\ra+\et^2 |f^{(2)}\ra+\dots,\no
\ti E(\et)=E+O(\et),
\nom}
where we take the finiteness condition for the energy $E(\et)$
in the limit $\et\to 0$ into account. In the lowest order in $\et$,
we have
 \disn{32}{
H^{(0)}|f^{(0)}\ra = 0.
\nom}
Hence,
 \disn{33}{
a(p)|f^{(0)}\ra=0,\quad p_-\le -\e.
\nom}
Further,
 \disn{34}{
H^{(1)}|f^{(0)}\ra+H^{(0)}|f^{(1)}\ra=0.
\nom}
By virtue of equality (\ref{32}), we hence have
 \disn{35}{
\langle f^{(0)}|H^{(1)}|f^{(0)}\ra =0
\nom}
or, equivalently,
 \disn{36}{
\langle f^{(0)}|\ls P_0+P_3\rs_{\La,x'^0=0}|f^{(0)}\ra=0
\nom}
We therefore conclude that the dependence of the state $|f^{(0)}\ra$
on the field modes $\f_{\La\et}$ and $\Pi_{\La\et}$ must correspond
to the minimum of the operator $\ls P_0+P_3\rs_{\La,x'^0=0}$.
Introducing the notation $|{\rm vac}_\La\ra$ for this
dependence, we obtain the general form of the basis for the states
$|f^{(0)}\ra$:
 \disn{37}{
\left\{|f^{(0)}\ra\right\}=\left\{
\prod_{n,p_n\ge\e}a^+(p_n)|0\ra\right\}\otimes
|{\rm vac}_\La\ra_{x'^0=0}.
\nom}
In the next order in $\et$, we have
 \disn{38}{
H^{(2)}|f^{(0)}\ra+H^{(1)}|f^{(1)}\ra+H^{(2)}|f^{(0)}\ra=E|f^{(0)}\ra.
\nom}
Hence, the values of $E$ are determined by the set
of eigenvalues of the operator
 \disn{39}{
\Pc_0 H^{(2)}\Pc_0=H_{LF}=
\int d^2 p_\p \int\limits_{p_-\ge\e}\!dp_-
\frac{m^2+p_\p^2}{2p_-}\,a^+(p)a(p)+\hfill\no+
\la\int d^2x^\p dx^-\ls
\fc_\e^4(x)+4\langle {\rm vac}_\La|\f_\La|{\rm vac}_\La\ra\fc_\e^3(x)+
6\langle {\rm vac}_\La|\f_\La^2|{\rm vac}_\La\ra\fc_\e^2(x)\rs
\str{\La\to\infty}\no\hfill\str{\La\to\infty}
H^{\rm can}_{LF}(\fc_\e)+
\la\int d^2x^\p d x^-\ls 4\langle\f\ra_{\rm vac}\fc^3_\e(x)+
6\langle\f^2\ra_{\rm vac}\fc^2_\e(x)\rs.
\nom}
where we let $\Pc_0$ denote the projection on the subspace of states
$|f^{(0)}\ra$ and $\langle\f^n\ra_{\rm vac}$ is the result of averaging
contributions of the corresponding field modes over $|{\rm vac}_\La\ra$.
In the framework of the described limiting transition, we cannot find
the constants $\langle\f\ra_{\rm vac}$ and $\langle\f^2\ra_{\rm vac}$,
which hence remain free parameters.

The last relation determines the desired approximate expression
for the Hamiltonian on the LF and its difference from the expression
$H^{\rm can}_{LF}$ obtained by directly quantizing on the LF.
We note that this result for the Hamiltonian on the LF is consistent
with the covariant perturbation theory in the coupling constant
in all orders \cite{tmf97}.

But the above method for constructing the Hamiltonian on the LF meets
difficulties in the gauge field case, where dynamical variables
are gauge-noninvariant fields. It is difficult to introduce
regularizations preserving Lorentz as well as gauge invariance.
Moreover, the proposed splitting of one part of the Fourier modes
of these fields from the other also breaks these symmetries if we
neglect intermediate modes. The presence of constraints between
canonical variables due to gauge symmetry complicates using the above
procedure to study the limiting transition on the LF because
these constraints are nonlinear in fields and the result may depend
on which variables are taken as independent (and are separated
into parts in passing to the Fourier modes).

\section{Gauge theory on the lattice}
We now restrict ourself to describing a gauge-invariant approach
for constructing Hamiltonians in the coordinates $\xt$,
which are close to the LF coordinates, as a preliminary step
to formulating and investigating the limiting transition on the LF
for the gauge field theory. This approach uses the space-time
lattice \cite{wils2} in the above coordinates and is analogous
to the method in \cite{creutz}.

We introduce the lattice in the coordinates $\xt^+$, $\xt^-$ and $\xt^\p$
with the parameters $a_+$, $a_-$ and $a_\p$ denoting the distances
between lattice sites along the corresponding coordinates in the scale
of those coordinates. Gauge fields are described by unitary $N\times N$
matrices $U_\m(\xt)$ from the group $SU(N)$.
These matrices are set into correspondence to the lattice links
as shown in the diagram:
 
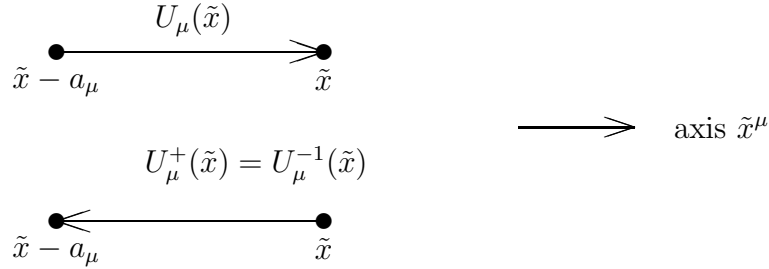
\begin{figure}[ht]
\unitlength 1.00mm
\linethickness{0.4pt}
\vskip -2mm
\begin{picture}(105.48,20)(-13,0)
\put(23.41,10.03){\line(1,0){35.56}}
\put(23.41,6.08){\makebox(0,0)[cc]{$\tilde x-a_\mu$}}
\put(58.97,6.38){\makebox(0,0)[cc]{$\tilde x$}}
\put(41.65,12.16){\makebox(0,0)[cb]{$U_\mu(\tilde x)$}}
\put(23.41,10.03){\circle*{2}}
\put(58.97,10.03){\circle*{2}}
\put(58.97,10.03){\line(-3,1){4.5}}
\put(58.97,10.03){\line(-3,-1){4.5}}
\put(84.81,0){\line(1,0){15.50}}
\put(100.31,0){\line(-3,1){4.0}}
\put(100.31,0){\line(-3,-1){4.0}}
\put(105.48,0){\makebox(0,0)[lc]{axis $\tilde x^\mu$}}
\end{picture}
\vskip 10mm
\begin{picture}(105.48,12.16)(-13,0)
\put(23.41,10.03){\line(1,0){35.56}}
\put(23.41,6.08){\makebox(0,0)[cc]{$\tilde x-a_\mu$}}
\put(58.97,6.38){\makebox(0,0)[cc]{$\tilde x$}}
\put(50,15){\makebox(0,0)[cb]{$U^+_\m(\tilde x)=U^{-1}_\mu(\tilde x)$}}
\put(23.41,10.03){\circle*{2}}
\put(58.97,10.03){\circle*{2}}
\put(23.41,10.03){\line(3,1){4.5}}
\put(23.41,10.03){\line(3,-1){4.5}}
\end{picture}
\vskip -2mm
\label{r1}
\end{figure}

Under the gauge transformations $\Om(\xt)$ corresponding
to the fundamental representation of the group $SU(N)$,
the matrices transform as
 \disn{40}{
U_\m(\xt)\stv{\Om}\Om(\xt)U_\m(\xt)\Om^{-1}(\xt-a_\m).
\nom}
The formal transition to the theory in the continuous space
with the gauge fields $\ti A_{\mu}(\ti x)=\ti A_{\mu}^a(\ti x)\lambda^a/2$,
where $\lambda^a/2$ are analogues of the Gell-Mann matrices for the group
$SU(N)$, corresponds to the representation
 \disn{41}{
U_\m(\xt)\approx e^{ia_\m \ti A_\m(\xt)}\approx 1+i a_\m \ti A_\m(\xt)+\dots
\nom}
under the condition $a_{\mu}\to 0$, where $\ti A_\m$ are the vectors
related to the coordinates $\xt$.

We also define the "plaquette" variables $U_{\mu\nu}(x)$:
 \disn{42}{
U_{\m\n}(\xt)=U_\m(\xt)U_\n(\xt-a_\m)U_\m^{-1}(\xt-a_\n)U^{-1}_\n(\xt),\no
U_{\m\n}(\xt)=U^+_{\n\m}(\xt)=U^{-1}_{\n\m}(\xt).
\nom}
Under the gauge transformations, we have
 \disn{43}{
U_{\m\n}(\xt)\stv{\Om}\Om(\xt)U_{\m\n}(\xt)\Om^{-1}(\xt),
\nom}
and the quantity ${\rm\bf tr} U_{\mu\nu}(x)$ is hence gauge invariant.
The analogue of formula (\ref{41}) for the plaquette variables is
 \disn{44}{
U_{\m\n}(\xt)\approx e^{ia_\m a_\n \ti F_{\m\n}(\xt)}\approx
1+ia_\m a_\n \ti F_{\m\n}(\xt)-\frac{1}{2}a^2_\m a^2_\n \ti F_{\m\n}^2(\xt)+\dots,
\nom}
where $\ti F_{\mu\nu}=\dd_\m \ti A_\n-\dd_n \ti A_\m-i[\ti A_\m,\ti A_\n]$.

Formulas (\ref{41}) and (\ref{44}) yield an expression approximating
the continuous-theory Lagrangian action in terms of lattice variables
using the prescriptions
 \disn{45}{
a_\m a_\n \ti F_{\m\n}(\xt)\lo \frac{1}{2i}\ls U_{\m\n}(\xt)-U^+_{\m\n}(\xt)\rs
\equiv \im U_{\m\n}(\xt),\no
a^2_\m a^2_\n\, {\rm\bf tr} \ti F^2_{\m\n}(\xt)\lo
{\rm\bf tr}\ls 2-U_{\m\n}(\xt)-U^+_{\m\n}(\xt)\rs
\equiv 2\,{\rm\bf tr}\ls 1-\re U_{\m\n}(\xt)\rs.
\nom}
In coordinates (\ref{12}) close to the LF, we have
 \disn{46}{
S=\int d^4\xt\, L(\xt),\no
L(\xt)=-\frac{1}{4g^2}\ti F^a_{\m\n}(\xt)\ti F^a_{\rho\la}(\xt)
\ti g^{\rho\m}(\xt)\ti g^{\la\n}(\xt)=\ns=
\frac{1}{2g^2}\ti F^a_{+-}(\xt)\ti F^a_{+-}(\xt)+
\sum_{k=1,2}\frac{1}{g^2}\ls
\frac{\et^2}{2}\ti F^a_{+k}(\xt)\ti F^a_{+k}(\xt)+
\ti F^a_{+k}(\xt)\ti F^a_{-k}(\xt)\rs+\ns
+\frac{1}{2g^2}\ti F^a_{12}(\xt)\ti F^a_{12}(\xt).
\nom}
in the continuous case. For the lattice action, we obtain
 \disn{47}{
S_{{\rm lat}}=\frac{2 a_+a_-a^2_\p}{g^2}\sum_x{\rm\bf tr}\,
\ls \frac{\re\ls 1-U_{+-}(\xt)\rs}{a^2_+a^2_-}+
\sum_{k=1,2}\frac{\et^2\re\ls 1-U_{+k}(\xt)\rs}{a_+^2a_\p^2}+
\right.\ns+\left.
\sum_{k=1,2}\frac{\im U_{+k}(\xt)\im U_{-k}(\xt)}{a_+a_-a_\p^2}-
\frac{\re\ls 1-U_{12}(\xt)\rs}{a_\p^4}\rs.
\nom}
To derive the analogue of the Hamiltonian, we consider the lattice
representation of the functional integral \cite{wils2} for the matrix
elements of the evolution operator \cite{creutz} (see \cite{knslav}
for an analogous consideration in
the continuous case). We use the coordinate $\xt^+$ as the evolution
parameter (subsequently considering the limit as $a_+\to 0$) and also
introduce the gauge condition
 \disn{47.1}{
U_+(\xt)=1.
\nom}
Under such a condition, it is easy \cite{creutz} to find the form
of the quantum evolution operator $\ti T_+$ relating the basis states
(the eigenvalues corresponding to these states are the integration
variables in the functional integral) at the time instants $\xt^+$
differing by the quantity $a_+$. The operator $\ti T_+$ is in turn
related to the Hamiltonian $\ti H$ in the coordinates $\xt$
in the limit as $a_+\to 0$:
 \disn{48}{
\ti T_+ =\exp\ls -i a_+\ti H+O(a_+^2)\rs.
\nom}

By analogy with \cite{creutz}, we introduce the basis of states
used when defining the functional integral,
 \disn{49}{
|U\ra= \prod_{i,\xt}|U_i(\xt)\ra,\quad i=-,1,2,\no
\hat U_i(\xt)|U\ra=U_i(\xt)|U\ra,\qquad
\langle U'|U\ra=\de(U',U),\qquad
1\equiv \int dU\, |U\ra\langle U|,
\nom}
where $dU$ is the invariant measure on the group of matrices $U$.

We define the unitary operators $\hat R_{i,\xt}(g_i)$
by analogy with the "shift" operation on the matrix group $U$:
 \disn{50}{
\hat R_{i,\xt}(g_i)|U_i(\xt)\ra=|g_iU_i(\xt)\ra,
\nom}
where $g_i$ is an arbitrary $N\times N$ $SU(N)$ matrix.

It is easy to verify that the operator $\ti T_+$
can be written as \cite{creutz}
 \disn{51}{
\ti T_+=\prod_{k}\prod_{\xt^-,\xt^\p}\int d g_- d g_\p \hat R_{-,\xt}(g_-)
\hat R_{k,\xt}(g_k)\times\ns\times
\exp\Biggl[
-\frac{2i}{g^2}{\rm\bf tr}\ls\frac{g_-+g_-^+}{2a_+a_-}\rs
-\frac{2 i\et^2 a_-}{g^2 a_+}{\rm\bf tr}\ls \frac{g_k+g_k^+}{2}\rs
-\ns-
\frac{2i}{g^2}{\rm\bf tr}\ls\frac{g_k-g_k^+}{2i}\im U_{-k}(\xt)+
\frac{2ia_+a_-}{g^2 a_\p^2}\re U_{12}(\xt)\rs\Biggr].
\nom}
We parameterize the matrices $g_i$ using the real parameters $\theta^a_i$:
 \disn{51.1}{
g_i=\exp\ls i \te_i^a\frac{\la^a}{2}\rs.
\nom}

For the operators $\hat R_{i,\xt}(g_i)$, we have
 \disn{52}{
\hat R_{i,\xt}(g_i)=\exp\ls i \te_i^a\hat \Pi^a_i(\xt)\rs
\nom}
where the operators $\hat \Pi^a_i(\xt)$ realize the representation
of the matrices $\lambda^a/2$ and satisfy the commutation relations
 \disn{53}{
\lks\hat\Pi^a_i(\xt),\hat\Pi^b_i(\xt)\rks=if^{abc}\hat\Pi^c_i(\xt)
\nom}
and $f^{abc}$ are the structure constants of the group $SU(N)$,
 \disn{54}{
\lks \frac{\la^a}{2},\frac{\la^b}{2}\rks=if^{abc}\frac{\la^c}{2}.
\nom}

In the limit as $a_+\to 0$, we can evaluate the integrals over $g_i$
in the expression for $\ti T_+$ in full analogy
with \cite{creutz}. As a result, we obtain the expression
for the Hamiltonian:
 \disn{55}{
\ti H(\et)=\sum_{\xt^-,x^\p}\Biggl[
\frac{g^2a_-}{2a^2_\p}\hat\Pi^a_-(\xt)\hat\Pi^a_-(\xt)+\ns+
\frac{g^2}{2\et^2 a_-}\sum_{k=1,2}\ls
\hat\Pi^a_k(\xt)-\frac{1}{g^2}{\rm\bf tr}\ls\la^a \im U_{-,k}(\xt)\rs\rs^2-
\frac{2a_-}{g^2 a^2_\p}{\rm\bf tr}\,\re U_{12}(\xt)\Biggr].
\nom}

By virtue of formulas (\ref{50}) and (\ref{52}), we have the relation
 \disn{56}{
e^{-i\te^a_i\hat\Pi^a_i}\hat U_i e^{i\te^a_i\hat\Pi^a_i}=
e^{i\te^a_i\la^a/2}\hat U_i,
\nom}
whence we can obtain the commutation relations for the operators
$\hat\Pi^a_i(\xt)$ and $\hat U_i(\xt)$:
 \disn{57}{
\lks\hat\Pi^a_i(\xt),\hat U_i(\xt)\rks=-\frac{\la^a}{2}\hat U_i(\xt).
\nom}
It is easy to verify that the form of these relations is invariant
under the gauge transformations of variables
if the quantity $\hat\Pi^a_i(x)$ obeys the transformation law
 \disn{58}{
\hat \Pi_i(\xt)\stv{\Om}\Om(\xt)\hat\Pi_i(\xt)\Om^{-1}(\xt),
\nom}
where $\hat\Pi_i(\xt) = (\lambda^a/2)\hat\Pi^a_i(\xt)$.
To obtain this result, it suffices to note that the matrices $\lambda^a$
are vectors in the adjoint representation of the group $SU(N)$.

Gauge condition (\ref{47.1}) preserves the symmetry under the gauge
transformations independent of $\xt^+$.
The generators of these transformations correspond to the canonical
constraints, which are regarded in the quantum theory as conditions
determining the physical subspace of states \cite{dirlek}. Vectors
of this subspace described by functionals of fields must be gauge
invariant. On a space lattice of finite size, these states correspond
to functions of traces of products of the matrices $U_i(\xt)$ along
all possible closed contours on the links of the lattice.
The Hamiltonian $\ti H(\eta)$ must be considered on only these states.
The physical vacuum must correspond to the minimum of the Hamiltonian.
Moreover, the vacuum state is assumed to be
invariant under shifts along the coordinates $\xt^-$ and $\xt^{\perp}$.
Analogously, when finding the spectrum of bound states, we can fix
the subspace of states that are invariant under shifts along $\xt^{\perp}$
and acquire a phase factor under shifts along $\xt^-$
(this corresponds to using Eqs.~(\ref{3}) in the continuous theory).

The exact solution of this problem is obviously as difficult as the
one in the standard Lorentzian coordinates with the Hamiltonian
at the fixed time $x^0$.

Passing to the coordinates $\xt$ close to the LF coordinates with
a small parameter $\eta$ allows using the smallness of $\eta$
if we make simple assumptions about a possible approximation
to the exact solution, as demonstrated above for the scalar field
theory in the continuous space. For example, we can develop the perturbation
theory in the parameter $\eta$ for the
equation for the eigenvalues of the Hamiltonian $\ti H(\eta)$ fixing
the lattice parameter values and spatial size (imposing periodic
boundary conditions in spatial variables on fields). Investigating
such a perturbation theory in the continuous space (with the fixed
cutoff parameter $|x^-|\le L$ and periodic boundary conditions
in the coordinate $x^-$) for two-dimensional quantum electrodynamics
demonstrated \cite{jf47,jf49,vest89,naus} that using
the phenomenologically fitted modification of the Hamiltonian terms
containing zero modes of the field, we can attain an adequate inclusion
of vacuum effects. This feeds the hope to find an analogous result
in the lattice approach.

On the other hand, using the perturbation theory in $\et$,
we can construct the lattice vacuum state described by the functional
of gauge fields, restricting ourself to just a few lowest orders in $\et$,
and then use this functional to calculate quantum correlation functions
approximately, preserving the finite, but small, value of the parameter $\et$.
Such an approach can be useful when analyzing hadron scattering at high
energies.

\vskip 1em
{\bf Acknowledgments.}
The authors thank the UNESCO Regional Bureau for Science and Culture
in Europe for supporting the V.~A.~Fock International School of Physics.
This work was supported in part (S.~A.~P. and E.~V.~P.) by the Russian
Foundation for Basic Research (Grant No.~05-02-17477) and the Russian
Federal Education Agency (Project No.~RNP.2.1.1.1112).

\end{document}